\documentclass{ifacconf}

\usepackage{graphicx}      
\usepackage{natbib}        
\usepackage{amsmath}
\usepackage{siunitx}
\usepackage{todonotes}

\usepackage{tikz}
\usetikzlibrary{arrows,automata,shapes}
\usepackage{xcolor}
\definecolor{green_node}{RGB}{188,213,190}
\definecolor{purple_label}{RGB}{157,153,173}

\usepackage{caption}
\usepackage{subcaption}
\usepackage{booktabs}
\usepackage{url}

\begin{document}
\begin{frontmatter}

\title{Ball in double hoop: demonstration model for numerical optimal control\thanksref{footnoteinfo}} 

\thanks[footnoteinfo]{This work was supported by the Grant Agency of the Czech Technical
University in Prague, grant No. SGS16/232/OHK3/3T/13, and partially by the Czech Science Foundation within the project P206/12/G014.}

\author{Martin Gurtner\quad} 
\author{\quad Ji\v{r}\'{i} Zem\'{a}nek} 

\address{Faculty of Electrical Engineering\\Czech Technical University in Prague\\Technicka 2, 166 27 Praha 6, Czech Republic\\\{martin.gurtner,jiri.zemanek\}@fel.cvut.cz}

\begin{abstract}                
Ball and hoop system is a well-known model for the education of linear control systems. In this paper, we have a look at this system from another perspective and show that it is also suitable for demonstration of more advanced control techniques. In contrast to the standard use, we describe the dynamics of the system at full length; in addition to the mode where the ball rolls on the (outer) hoop we also consider the mode where the ball drops out of the hoop and enters a free-fall mode. Furthermore, we add another (inner) hoop in the center upon which the ball can land from the free-fall mode. This constitutes another mode of the hybrid description of the system. We present two challenging tasks for this model and show how they can be solved by trajectory generation and stabilization. We also describe how such a model can be built and experimentally verify the validity of our approach solving the proposed~tasks.\footnote{All codes and drawings are available at \url{http://github.com/aa4cc/flying-ball-in-hoop}}
\end{abstract}


\end{frontmatter}

\section{Introduction} 
Ball and hoop system is well-established among laboratory models used in teaching control. Its origin dates back to~\cite{Wellstead1980Scale,Wellstead1983Ball} who introduced it as a simple model having qualitatively the same dynamics as the liquid slop problem; in addition, \cite{Wellstead1980Scale,Wellstead1983Ball} and \cite{Fabregas2011Developing} also described how this model can be used in teaching of linear control theory. More specifically, the model was used in a regime where the ball is close to the stable position on the hoop and where a linear approximation of motion dynamics is valid.

In contrast, we consider full repertoire of motions the system offers. We describe the system as a hybrid model with a mode where the ball rolls on the outer hoop and with a mode where the ball has dropped out and entered a free-fall mode. This description enables us to generate dynamically richer trajectories (for instance, the ball can go through the top of the hoop or even fly in free fall); nevertheless, such trajectories also require more advanced control techniques. First, one needs to find a way to generate such trajectories. Second, the trajectories have to be stabilized; in other words, one cannot simply take the controls corresponding to a desired trajectory, apply it to the system and expect to observe the desired trajectory. This model is ideal for a demonstration of how these techniques, trajectory generation and stabilization, can be applied because it is relatively simple and allows one to pose tasks which are rather difficult to solve by other means. We propose and solve two such tasks. Specifically, we formulate each task as an \textit{Optimal Control Problem (OCP)} with nonlinear dynamics and limited control input. We solve the OCPs and show how to stabilize the obtained trajectories.

In addition, we added another hoop inside the outer one. This further extends the variety of tasks that can be solved on this system; one can, for example, swing up the ball on the outer hoop, let it fall and stabilize it on the inner hoop. It is worthy of note that the stabilization of the ball on the inner hoop can be viewed as stabilization of a disk on a disk; that itself is an interesting problem (for references, see~\cite{Ryu2013Control}).

This paper is organized as follows. In Section 2, we derive a mathematical model of the system. In the following section, we present two examples of interesting and challenging tasks for this system and propose an approach solving them. In Section 4, we describe hardware setup of the system and verify the viability of the proposed approach solving the tasks.


\section{Modeling} 
\label{sec:model}
\begin{figure}[t]
\begin{center}
\includegraphics[width=5.5cm]{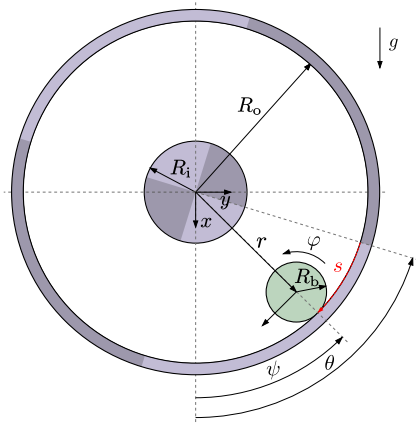}    
\caption{A sketch of the ball and hoop system.}
\label{fig:model_sketch}
\end{center}
\end{figure}

\begin{figure}[t]
  \center
  \begin{tikzpicture}[->,>=stealth',shorten >=1pt,auto,node distance=2.8cm,
                      semithick,scale=0.9, every node/.style={scale=0.9}]
    \tikzstyle{every state}=[fill=green_node,draw=none,text=black]
  
    \node[state,ellipse]    (A)                   {Outer hoop};
    \node[state,ellipse]    (B)[right of=A]       {Free fall};
    \node[state,ellipse]    (C)[right of=B]       {Inner hoop};
    \node[circle,inner sep=1.2pt,fill=purple_label]                   (A_label)[below of=A,node distance=0.8cm,text=white]       {S1};
    \node[circle,inner sep=1.2pt,fill=purple_label]                   (B_label)[below of=B,node distance=0.8cm,text=white]       {S2};
    \node[circle,inner sep=1.2pt,fill=purple_label]                   (C_label)[below of=C,node distance=0.8cm,text=white]       {S3};
  
    \path (A) edge  [bend left]       node[pos=0.45] {$\gamma_1>0$} (B)
        (B) edge  [bend left]        node[pos=0.55] {$\gamma_2>0$} (A)

        (B) edge  [bend left]       node[pos=0.55] {$\gamma_3>0$} (C)
        (C) edge  [bend left]        node[pos=0.45] {$\gamma_4>0$} (B);
          
  \end{tikzpicture}
  \caption{Modes of the hybrid description of the ball and hoop system.}
  \label{fig:hybrid_modes}
\end{figure}
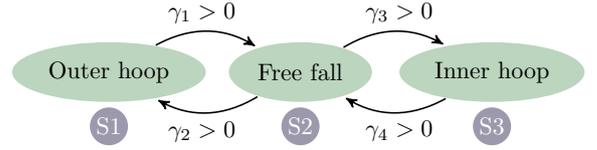

The system is cartooned in Fig.~\ref{fig:model_sketch} and the modes of the hybrid description are displayed in Fig.~\ref{fig:hybrid_modes}. The hybrid description has three modes: the ball rolls on the outer hoop (S1), the ball is in free fall (S2), and the ball rolls on the inner hoop (S3).

Before we delve into the derivation of equations of motion for each mode, let us devote a few words to coordinate systems in which we will describe the motion of the ball. Due to the rotational symmetry, polar coordinates are a natural choice for modes S1 and S3. In contrast, Cartesian coordinates are more suitable for the free-fall mode (S2) because then the equations of motion are linear. In fact, it turns out that the description of motion in S2 in polar coordinates is strongly nonlinear. Thus, from the implementation point of view (i.e. to increase numerical stability), one should use polar coordinates for modes S1 and S3 and Cartesian coordinates for S2. Nevertheless, transformations of the coordinates would make equations in this paper unnecessarily complicated, and thus we stick to polar coordinates for all three modes.

For simplicity, we will not state the time dependence explicitly. At first, we derive equations of motion for each mode and after that, we describe guards and reset maps for transitions between the modes.

\subsection{Outer hoop (S1)} 
\label{sub:outer_hoop}
A model of a ball rolling in a hoop was derived by~\cite{Wellstead1983Ball}. For reader's convenience---and because Wellstead made a small mistake in the derivation of the model---we rederive the model.

We use \textit{Euler-Lagrange equation} to derive the model. Let us choose angles $\psi$ and $\theta$ (see Fig.~\ref{fig:model_sketch}) as the generalized coordinates. We assume that the ball rolls without slipping. In addition, we do not consider the dynamics of the hoop as we assume that we can directly command the angular acceleration of the hoop $\ddot{\theta}$. In fact, the acceleration will be our input to the model and it constitutes a rheonomic (time-varying) constraint. This constraint reduces the number of independent coordinates to one.

The kinetic co-energy of the ball is
\begin{equation}
  T^\ast = \frac{1}{2} mv^2 + \frac{1}{2}I(\dot{\varphi} + \dot{\theta})^2,
\end{equation}
where $m$ is the mass of the ball, $I$ is the moment of inertia of the ball and $\varphi=\frac{s}{R_\mathrm{b}}=\frac{R_\mathrm{o}}{R_\mathrm{b}}(\theta-\psi)$. The translational velocity $v$ and the angular velocity $\dot{\varphi}$ of the ball can be expressed as follows
\begin{subequations}
\label{eq:kinematicConstraints}
  \begin{align}
  \label{eq:kinematicConstraints_vt}
    v &= -\left(R_\mathrm{o}-R_\mathrm{b}\right)\dot{\psi}, \\
  \label{eq:kinematicConstraints_phi}
    \dot{\varphi} &= \frac{R_\mathrm{o}}{R_\mathrm{b}}\left(\dot{\theta} - \dot{\psi}\right).  
  \end{align}
\end{subequations}
Thus, the kinetic energy expressed in the generalized coordinates is 
\begin{equation}
  T^\ast = \frac{1}{2} m\left(R_\mathrm{o}-R_\mathrm{b}\right)^2\dot{\psi}^2
  +
  \frac{1}{2}I\left(\frac{R_\mathrm{o}+R_\mathrm{b}}{R_\mathrm{b}}\dot{\theta} - \frac{R_\mathrm{o}}{R_\mathrm{b}}\dot{\psi}\right)^2.
\end{equation}
Potential energy of the ball is given by
\begin{equation}
  V = -mg\left(R_\mathrm{o}-R_\mathrm{b}\right)\cos\psi
\end{equation}
and the system content modeling friction of the ball is
\begin{equation}
  D = \frac{1}{2}b\dot{\varphi}^2=\frac{1}{2}b\frac{R_\mathrm{o}^2}{R_\mathrm{b}^2}\left(\dot{\theta} - \dot{\psi}\right)^2.
\end{equation}

Defining the Lagrangian $\mathcal{L}=T^\ast-V$, an equation of motion of the ball can be computed by Euler-Lagrange equation as
\begin{equation}
  \frac{\mathrm{d}}{\mathrm{d}t}\left( \frac{\partial \mathcal{L}}{\partial \dot{\psi}} \right)
  -
  \frac{\partial \mathcal{L}}{\partial \psi}
  +
  \frac{\partial D}{\partial \dot{\psi}}
  =
  0
\end{equation}
which results in the following differential equation
\begin{equation}
  \bar{a}_\mathrm{o}\ddot{\psi} + \bar{b}_\mathrm{o}\dot{\psi} + \bar{c}_\mathrm{o}\sin\psi + \bar{d}_\mathrm{o}\dot{\theta} = \bar{e}_\mathrm{o}\ddot{\theta},
\end{equation}
with coefficients
\begin{subequations}
  \begin{align}
    \bar{a}_\mathrm{o} &= m \left(R_\mathrm{o}-R_\mathrm{b}\right)^2 + I \frac{R_\mathrm{o}^2}{R_\mathrm{b}^2}, \\
    \bar{b}_\mathrm{o} &= b  \frac{R_\mathrm{o}^2}{R_\mathrm{b}^2}, \\
    \bar{c}_\mathrm{o} &= m g \left(R_\mathrm{o}-R_\mathrm{b}\right), \\
    \bar{d}_\mathrm{o} &= -\bar{b}_\mathrm{o}, \\
    \bar{e}_\mathrm{o} &= I \frac{R_\mathrm{o}}{R_\mathrm{b}}\left( \frac{R_\mathrm{o}}{R_\mathrm{b}} + 1 \right).
  \end{align}
\end{subequations}

\subsection{Free fall (S2)} 
\label{sub:free_fall}
The motion of the center of the ball in free fall can be easily described in Cartesian coordinates because the only force acting on the ball is the gravitational force. Considering the Cartesian coordinate system shown in Fig.~\ref{fig:model_sketch}, we have 
\begin{equation}
  \ddot{x} = g
  \qquad \mathrm{and} \qquad
  \ddot{y} = 0.
\end{equation}
In order to get equations of motion in polar coordinates $(r,\psi)$ (see Fig.~\ref{fig:model_sketch}), we only need to transform the coordinates by relations $x=r\cos\psi$ and $y=r\sin\psi$. This way, we get
\begin{subequations}
  \begin{align}
    \ddot{r} &= r\dot{\psi}^2 + g\cos\psi, \\
    \ddot{\psi} &= -\frac{1}{r}\left(g\sin\psi + 2\dot{\psi}\dot{r}\right).
  \end{align}
\end{subequations}

To fully describe the state of the system during free fall of the ball we also need to describe evolution of $\varphi$; we assume that the angular velocity $\dot{\varphi}$ is constant during the free fall.

\begin{figure}
\begin{center}
\includegraphics[width=3.7cm]{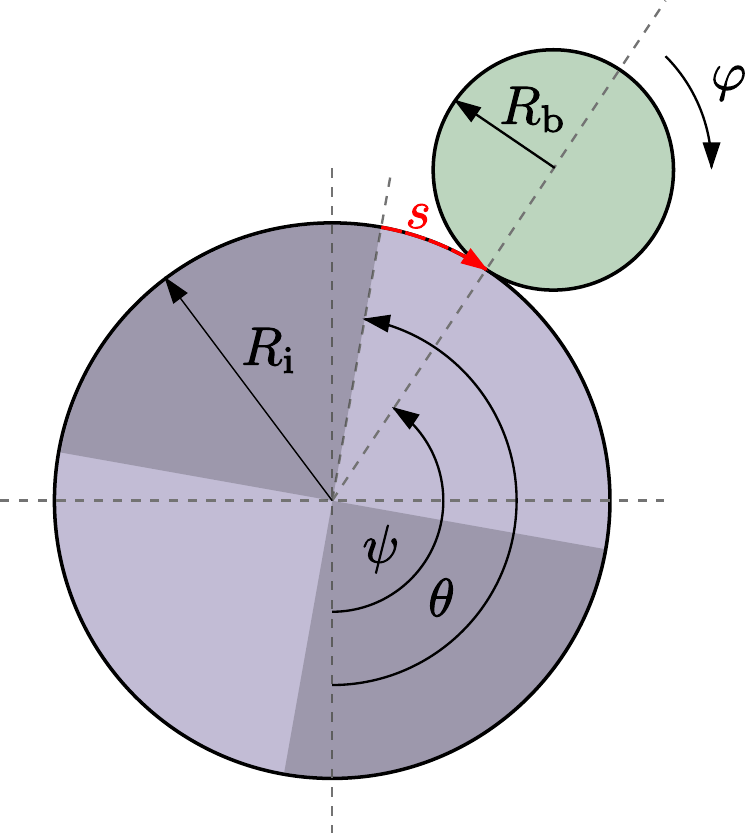}    
\caption{A sketch of the ball rolling on the inner hoop.}
\label{fig:model_sketch_inner}
\end{center}
\end{figure}

\subsection{Inner hoop (S3)} 
\label{sub:inner_hoop}
A sketch of the ball rolling on the inner hoop is shown in Fig.~\ref{fig:model_sketch_inner}. The equations of motion for the ball rolling on the inner hoop can be derived along similar lines as for the ball on the outer hoop. The only difference is that the effective rolling radius is $R_\mathrm{i}+R_\mathrm{b}$ instead of $R_\mathrm{o}-R_\mathrm{b}$ and angle $\varphi$ has the opposite orientation. Therefore, we skip the derivation and state directly the resulting equations of motion:
\begin{equation}
  \bar{a}_\mathrm{i}\ddot{\psi} + \bar{b}_\mathrm{i}\dot{\psi} + \bar{c}_\mathrm{i}\sin\psi + \bar{d}_\mathrm{i}\dot{\theta} = \bar{e}_\mathrm{i}\ddot{\theta},
\end{equation}
where
\begin{subequations}
  \begin{align}
    \bar{a}_\mathrm{i} &= m \left(R_\mathrm{i}+R_\mathrm{b}\right)^2 + I \frac{R_\mathrm{i}^2}{R_\mathrm{b}^2}, \\
    \bar{b}_\mathrm{i} &= b  \frac{R_\mathrm{i}^2}{R_\mathrm{b}^2}, \\
    \bar{c}_\mathrm{i} &= m g \left(R_\mathrm{i}+R_\mathrm{b}\right), \\
    \bar{d}_\mathrm{i} &= -\bar{b}_\mathrm{i}, \\
    \bar{e}_\mathrm{i} &= I \frac{R_\mathrm{i}}{R_\mathrm{b}}\left( \frac{R_\mathrm{i}}{R_\mathrm{b}} - 1 \right).
  \end{align}
\end{subequations}

\subsection{Transition between modes} 
\label{sub:transition_between_the_modes}

\subsubsection{Transition from S1 to S2} 
Mode S1 (outer hoop) is valid as long as the centripetal force $F_\mathrm{c}$ acting on the ball is larger then the normal component of the gravitational force $F_\mathrm{g}$ with respect to the hoop. Mathematically, it is valid if
\begin{equation}
\label{eq:mode1_validCondition}
  F_\mathrm{g}\cos\psi + F_\mathrm{c} = m g \cos\psi +  m\left( R_\mathrm{o} - R_\mathrm{b}\right)\dot{\psi}^2 > 0.
\end{equation}
If this condition does not hold, the ball drops of the hoop. It readily follows from~\eqref{eq:mode1_validCondition} that the guard for transition from S1 to S2 is
\begin{equation}
  \gamma_1  = - g \cos \psi -  \left( R_\mathrm{o} - R_\mathrm{b}\right)\dot{\psi}^2 > 0.
\end{equation}

The reset map for states $r$, $\dot{r}$, $\varphi$ and $\dot{\varphi}$ is
\begin{subequations}
\label{eq:resetMapS1_to_S2}
\begin{align}
  r^+            &= R_\mathrm{o}-R_\mathrm{b}, \\
  \dot{r}^+      &=   0, \\
  \varphi^+      &=   (\theta^- - \psi^-)\frac{R_\mathrm{o}}{R_\mathrm{b}}, \\
\label{eq:resetMapS1_to_S2_Dphi}  \dot{\varphi}^+  &= \frac{R_\mathrm{o}+R_\mathrm{b}}{R_\mathrm{b}}\dot{\theta} - \frac{R_\mathrm{o}}{R_\mathrm{b}}\dot{\psi}.
\end{align}
\end{subequations}
The first three relations are apparent. The last one is derived from the inertial angular velocity $(\dot{\varphi}^- + \dot{\theta}^-)$ of the ball and~\eqref{eq:kinematicConstraints_phi}. The remaining states transit to the mode S2 without any change, that is $\theta^+=\theta^-$, $\dot{\theta}^+=\dot{\theta}^-$, $\psi^+=\psi^-$ and $\dot{\psi}^+=\dot{\psi}^-$.

\subsubsection{Transition from S2 to S1} 
\label{ssub:transition_from_s2_to_s1}
This transition occurs when the ball hits the outer hoop. Thus the guard is
\begin{equation}
  \gamma_2 = r-R_\mathrm{o} + R_\mathrm{b}  > 0
\end{equation}
and the corresponding reset map for $\dot{\psi}$ is
\begin{equation}
\label{eq:resetMapS2_to_S1}
      \dot{\psi}^+ = \dot{\psi}^- + \dot{\psi}_\mathrm{rot}^-
\end{equation}
where $\dot{\psi}_\mathrm{rot}^-$ is derived from the angular velocity of the ball $\dot{\varphi}^-$ and can be computed from~\eqref{eq:kinematicConstraints_phi}:
\begin{equation}
      \dot{\psi}_\mathrm{rot}^- = \dot{\theta}^- - \frac{R_\mathrm{b}}{R_\mathrm{o}}\dot{\varphi}^-.
\end{equation}
The remaining states transit unchanged, that is $\theta^+=\theta^-$, $\dot{\theta}^+=\dot{\theta}^-$ and $\psi^+=\psi^-$.

\subsubsection{Transition from S2 to S3} 
\label{ssub:transition_from_s2_to_s3}
Analogously to the previous case, the transit occurs when the ball hits the inner hoop and the guard is
\begin{equation}
  \gamma_3 = R_\mathrm{i} + R_\mathrm{b} - r > 0.
\end{equation}
The reset map is also almost the same; the only difference is that the angular velocity $\dot{\psi}_\mathrm{rot}^-$ here is computed as follows
\begin{equation}
\label{eq:resetMapS3_to_S2_Dpsi}
    \dot{\psi}_\mathrm{rot}^- = \dot{\theta}^- + \frac{R_\mathrm{b}}{R_\mathrm{i}}\dot{\varphi}^-.
\end{equation}

\subsubsection{Transition from S3 to S2} 
Similarly to the transition from the outer hoop to free fall, the transition from the inner hoop to free fall occurs when the centripetal force becomes larger then the normal component of the gravitational force with respect to the hoop, that is
$$
  \gamma_4 = g\cos{\psi} + \left(R_\mathrm{i} + R_\mathrm{b}\right)\dot{\psi}^2 > 0.
$$
and analogously to~\eqref{eq:resetMapS1_to_S2}, the reset map is
\begin{subequations}
\label{eq:resetMapS3_to_S2}
\begin{align}
  r^+            &= R_\mathrm{i}+R_\mathrm{b}, \\
  \dot{r}^+      &=   0, \\
  \varphi^+      &=   -(\theta^- - \psi^-)\frac{R_\mathrm{i}}{R_\mathrm{b}}, \\
  \dot{\varphi}^+  &=   -\left(\frac{R_\mathrm{i}-R_\mathrm{b}}{R_\mathrm{b}}\dot{\theta} - \frac{R_\mathrm{i}}{R_\mathrm{b}}\dot{\psi}\right),
\end{align}
\end{subequations}
with the remaining states $\theta$, $\dot{\theta}$, $\psi$ and $\dot{\psi}$ transiting unchanged.

\section{Examples of interesting tasks} 
\label{sec:interesting_task_to_solve}
The hybrid description of the system allows us to pose visually appealing and from control perspective challenging tasks. In this section, we show two such tasks and describe control algorithms solving them.

\begin{figure}
\begin{center}
\includegraphics[width=8.4cm]{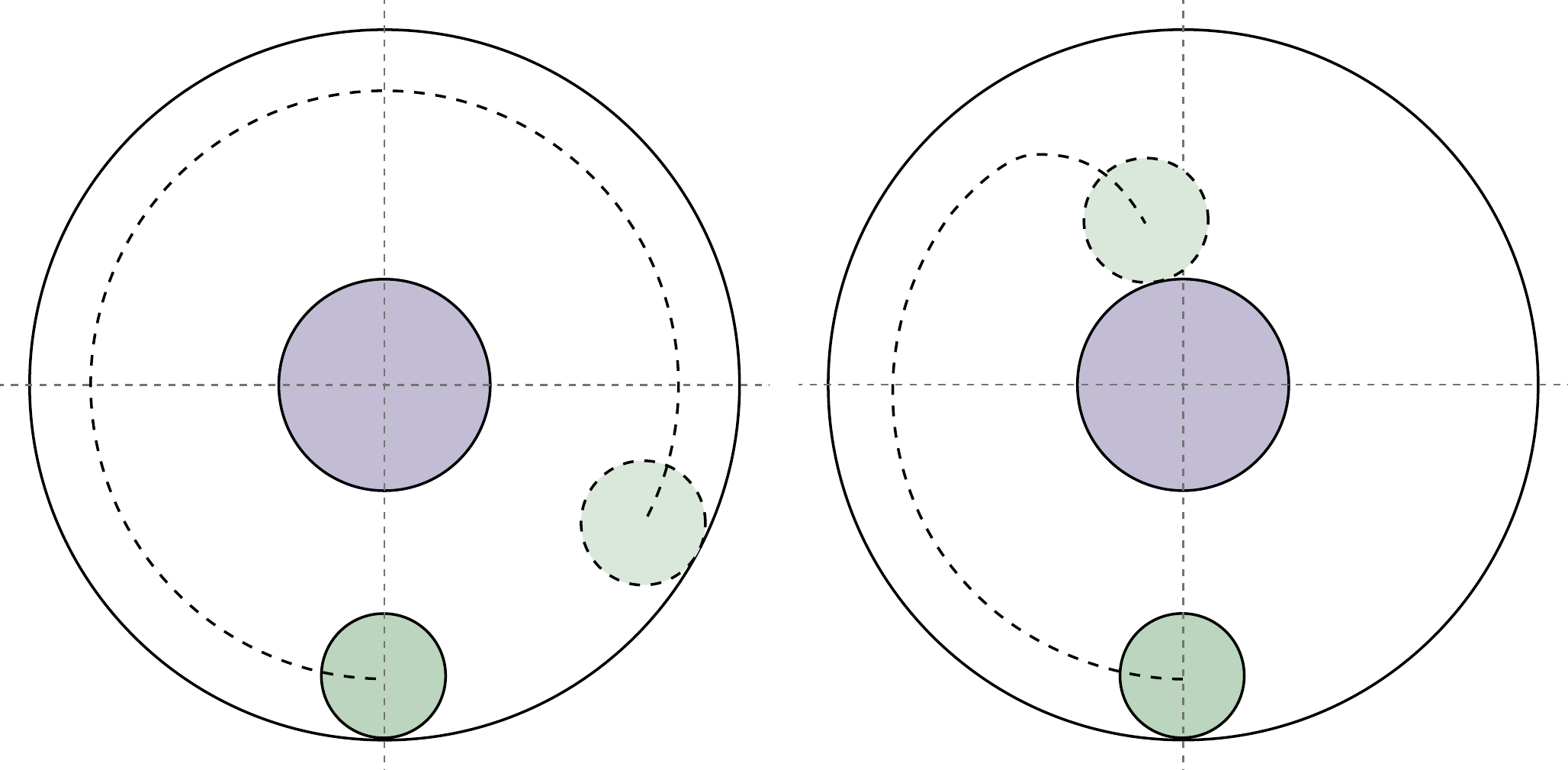}    
\caption{A cartoon of the challenging tasks to be solved with the ball and hoop system.}
\label{fig:challenging_tasks}
\end{center}
\end{figure}

\subsection*{Task 1: Roll the ball around the outer hoop} 
\label{sub:task_1_roll_the_ball_around_the_outer_hoop}
This task is sketched in Fig.~\ref{fig:challenging_tasks} on the left side. In the beginning, the ball is in the steady position on the outer hoop and the goal is to roll it around the hoop and get it back to the steady position. One might roughly imagine this task as the famous \textit{Loop-the-Loop} stunt where a car rides around a vertical circle.

Naturally, this task is an instance of an \textit{Optimal Control Problem (OCP)}. The ball is supposed to stay on the outer hoop and thus the hybrid model remains in the mode S1. To simplify the notation, let us define a state vector as
\begin{equation}
\label{eq:state_vectorS1_S3}
  \mathbf{x} = \left[\theta, \dot{\theta}, \psi, \dot{\psi} \right]^\top.
\end{equation}
and the state-space description corresponding to the mode S1 as 
\begin{equation}
\dot{\mathbf{x}} = f_\mathrm{S1}(\mathbf{x}, u),
\end{equation}
where $u = \ddot{\theta}$ and
\begin{equation}
  f_\mathrm{S1}(\mathbf{x})
  =
  \begin{bmatrix}
    x_2 \\
    u \\
    x_4 \\
    \frac{1}{\bar{a}_\mathrm{o}}  \left(-\bar{b}_\mathrm{o} x_4 - \bar{c}_\mathrm{o} \sin x_3 - \bar{d}_\mathrm{o} x_2 + \bar{e}_\mathrm{o} u\right)
  \end{bmatrix}.
\end{equation}

Now, we formulate the task as an OCP. In the initial time $T_0:=0$, everything is steady and thus the initial condition is $\mathbf{x}(0)=[0,0,0,0]^\top$. In a final time $T_\mathrm{f}$, we require that the ball has rolled around the hoop, is back in the steady state and the hoop does not move; that imposes conditions $x_3(T_\mathrm{f}) = -2\pi$, $x_4(T_\mathrm{f}) = 0$ and $x_2(T_\mathrm{f}) = 0$. Furthermore, to ensure that the ball does not drop out of the hoop, we require that the guard $\gamma_1$ is inactive during the whole trajectory. Instead of specifying the final time, we leave it as an optimization variable; we only require that the final time is smaller than a certain value $T_\mathrm{max}$. We also put bounds on the control $u$. Finally, the objective is to minimize the control $u$. Altogether, the OCP is
\begin{align}
\label{eq:rollIt_OCP}
  \min_{u(t), \mathbf{x}(t),T_\mathrm{f}} &\,\,\,  \int_{0}^{T_\mathrm{f}} u^2(t) \mathrm{d}t, \\
 \nonumber       \text{subject to:}\,     &\,\,\, x(0) = [0,0,0,0]^\top, \\
 \nonumber                          &\,\,\, x_2(T_\mathrm{f}) = 0, \\
 \nonumber                          &\,\,\, x_3(T_\mathrm{f}) = -2\pi, \\
 \nonumber                          &\,\,\, x_4(T_\mathrm{f}) = 0, \\
 \nonumber                          &\,\,\, \gamma_1\left(\mathbf{}{x}(t)\right) < 0, \\
 \nonumber                          &\,\,\, 0 < T_\mathrm{f} < T_\mathrm{max}, \\
 \nonumber                          &\,\,\, |u(t)| \leq u_\mathrm{max}, \\
 \nonumber                          &\,\,\, \dot{\mathbf{x}}(t) = f_\mathrm{S1}(\mathbf{x}(t), u(t)).
\end{align}

To solve this OCP we use \textit{direct collocation}~(\cite{Hargraves1987Direct}) to discretize it to $N$ equidistantly distributed knots in time and formulate it as a \textit{nonlinear programming problem (NLP)}. This simplification, however, comes at a cost. NLP solvers usually does not guarantee optimality and provide only suboptimal solutions. Let us denote a (possibly suboptimal) solution provided by a NLP solver by $u^\ast$, $\mathbf{x}^\ast$ and $T_\mathrm{f}^\ast$.

By solving the NLP, we obtain a discrete trajectory of the system evaluated at certain time instants separated by sampling time $T_\mathrm{s}=\frac{T_\mathrm{f}}{N}$. Ideally, if the model captures the real system well enough and the sampling time $T_\mathrm{s}$ is short enough (say, a few milliseconds), an application of the control input $u^\ast(t)$ results in the desired state trajectory $\mathbf{x}^\ast$. Nevertheless, short sampling periods imply large number of knots $N$ and thus make the NLP larger and possibly intractable. Furthermore, the model usually mismatches the real system to some extent. Therefore, we need to stabilize the trajectory; we need to keep the system as close as possible to the desired trajectory.

\subsubsection{Trajectory stabilization} 
\label{ssub:trajectory_stabilization}
To stabilize the trajectory, we invoke the powerful framework of \textit{Linear Quadratic Regulators (LQR)} and \textit{neighboring extremals}~(\cite{Bryson1975Applied}). We linearize the model around the trajectory and thus obtain a linear time-varying deviation model
\begin{equation}
  \delta\dot{\mathbf{x}}(t) = \mathbf{A}(t)\delta\mathbf{x}(t) + \mathbf{B}(t)\delta u(t), 
\end{equation}
where $\delta\mathbf{x} = \mathbf{x} - \mathbf{x}^\ast$ and $\delta u = u - u^\ast$. The time-varying matrices are
\begin{align}
  \mathbf{A}(t) &= \frac{\partial f_\mathrm{S1} \left(\mathbf{x}(t), u(t)\right)}{\partial \mathbf{x}}\Big|_{\mathbf{x}(t) = \mathbf{x}^\ast(t), u(t)=u^\ast(t)}, \\
  \mathbf{B}(t) &= \frac{\partial f_\mathrm{S1} \left(\mathbf{x}(t), u(t)\right)}{\partial u}\Big|_{\mathbf{x}(t) = \mathbf{x}^\ast(t), u(t)=u^\ast(t)}.
\end{align}
Now we design an LQR stabilizing the deviation model hence keeping $\delta\mathbf{x}(t)$
small and the system close to $\mathbf{x}^\ast$. We stick to a common notation and denote the state and input weight matrices by $\mathbf{Q}$ and $\mathbf{R}$, respectively. The control law of the LQR is
\begin{equation}
  \delta u(t) = -\mathbf{R}^{-1}\mathbf{B}^\top(t)\mathbf{S}(t) \delta \mathbf{x}(t),
\end{equation}
where matrix $\mathbf{S}(t)$ is given by the solution of the differential Riccati equation
\begin{equation}
  -\dot{\mathbf{S}}(t) = \mathbf{S}(t)\mathbf{A}(t) + \mathbf{A}\!\!^\top(t)\mathbf{S}(t) - \mathbf{S}(t)\mathbf{B}(t)\mathbf{R}^{-1}\mathbf{B}\!^\top(t)\mathbf{S}(t) + \mathbf{Q} 
\end{equation}
with final time condition $\mathbf{S}(T_\mathrm{f}) = \mathbf{Q}$.

Finally, the applied control to the system is given by a sum of the nominal input $u^\ast(t)$ and the trajectory stabilizing term $\delta u(t)$:
\begin{equation}
\label{eq:ctrl_law}
  u(t) = u^\ast(t) + \delta u(t).
\end{equation}


\subsection*{Task 2: Get the ball on the inner hoop} 
\label{sub:get_the_ball_on_the_inner_hoop}
The second task we would like to solve is to get the ball from the steady position on the outer hoop to the top of inner hoop and stabilize it there. This task is sketched in Fig.~\ref{fig:challenging_tasks} on the right side. Once again, we formulate this problem as an OCP, find a trajectory solving the problem and design a regulator stabilizing this trajectory.

Apparently, the trajectory we are looking for starts and ends in different modes of the hybrid model. This poses especially difficult problem to the trajectory generation because in the discretization of the corresponding OCP, one does not know how many knots should be assigned to each mode. Fortunately, the requirement that the ball ends up on the inner hoop can be reformulated to a requirement that the ball ends up on the outer hoop at specific angle $\psi_\mathrm{des}$ with specific angular velocity $\dot{\psi}_\mathrm{des}$. This is due to the fact that the free-fall mode is uncontrolled and thus the state of the system leaving the free-fall mode is fully determined by the state of the system entering the free-fall mode. Therefore, we can find a trajectory solving Task 2 by the same approach as we used for Task 1.

In addition to $\psi_\mathrm{des}$ and $\dot{\psi}_\mathrm{des}$, we also require that the hoop has specific angular velocity $\dot{\theta}_\mathrm{des}$ at the moment when the ball drops out of it. That is because by reset map~\eqref{eq:resetMapS1_to_S2_Dphi}, $\dot{\theta}_\mathrm{des}$ determines $\dot{\varphi}$ during the free fall and thus by reset map~\eqref{eq:resetMapS3_to_S2_Dpsi} also $\dot{\psi}$ just after the ball lands on the inner hoop. Therefore, we compute $\dot{\theta}_\mathrm{des}$ so that $\dot{\psi}$ after the impact is minimized.

The OCP for Task 2 has exactly the same form as~\eqref{eq:rollIt_OCP}; the only difference here is that the conditions for the final states are
\begin{subequations}
  \begin{align}
   x_2(T_\mathrm{f}) = \dot{\theta}_\mathrm{des}, \\
   x_3(T_\mathrm{f}) = \psi_\mathrm{des}, \\
   x_4(T_\mathrm{f}) = \dot{\psi}_\mathrm{des}.
  \end{align}
\end{subequations}
The remaining steps to solve this task are almost the same as with the previous one. The only additional step is to design a regulator stabilizing the ball on the inner hoop. After the ball lands, we hand over the control to a stationary LQR designed for this purpose.

\section{Experimental verification} 
\label{ssub:experimental_verification}
To verify the validity of our approach, we implemented the proposed control algorithm and tested it on a real hardware setup.

\begin{table}[t]
\begin{center}
\captionsetup{width=.9\columnwidth}
\caption{Parameters of the demonstration model}
\label{tb:demmodel_params}
\begin{tabular}{lccc}
Description & Parameter & Value & Unit \\\hline
Radius of the outer hoop & $R_\mathrm{o}$ & 95.8 & \si{\milli\meter}  \\
Radius of the inner hoop & $R_\mathrm{i}$ & 43.8 & \si{\milli\meter}  \\
Radius of the ball & $R_\mathrm{b}$ & 7.7 & \si{\milli\meter}  \\
Inertia of the ball & $I$ & $1.28\cdot10^{-6}$ & \si{\kilo\gram\meter\squared}  \\
Mass of the ball & $m$ & 0.032 & \si{\kilo\gram}  \\
Friction of the ball & $b$ & $1.4\cdot10^{-6}$ & \si{\newton\meter\second}  \\
Gravitational constant & $g$ & $9.81$ & \si{\meter\per\second\squared}  \\
\hline
\end{tabular}
\end{center}
\end{table}

\subsection{Construction of the model} 
\label{sec:construction_of_the_model}
\begin{figure}
\begin{center}
\includegraphics[width=6cm]{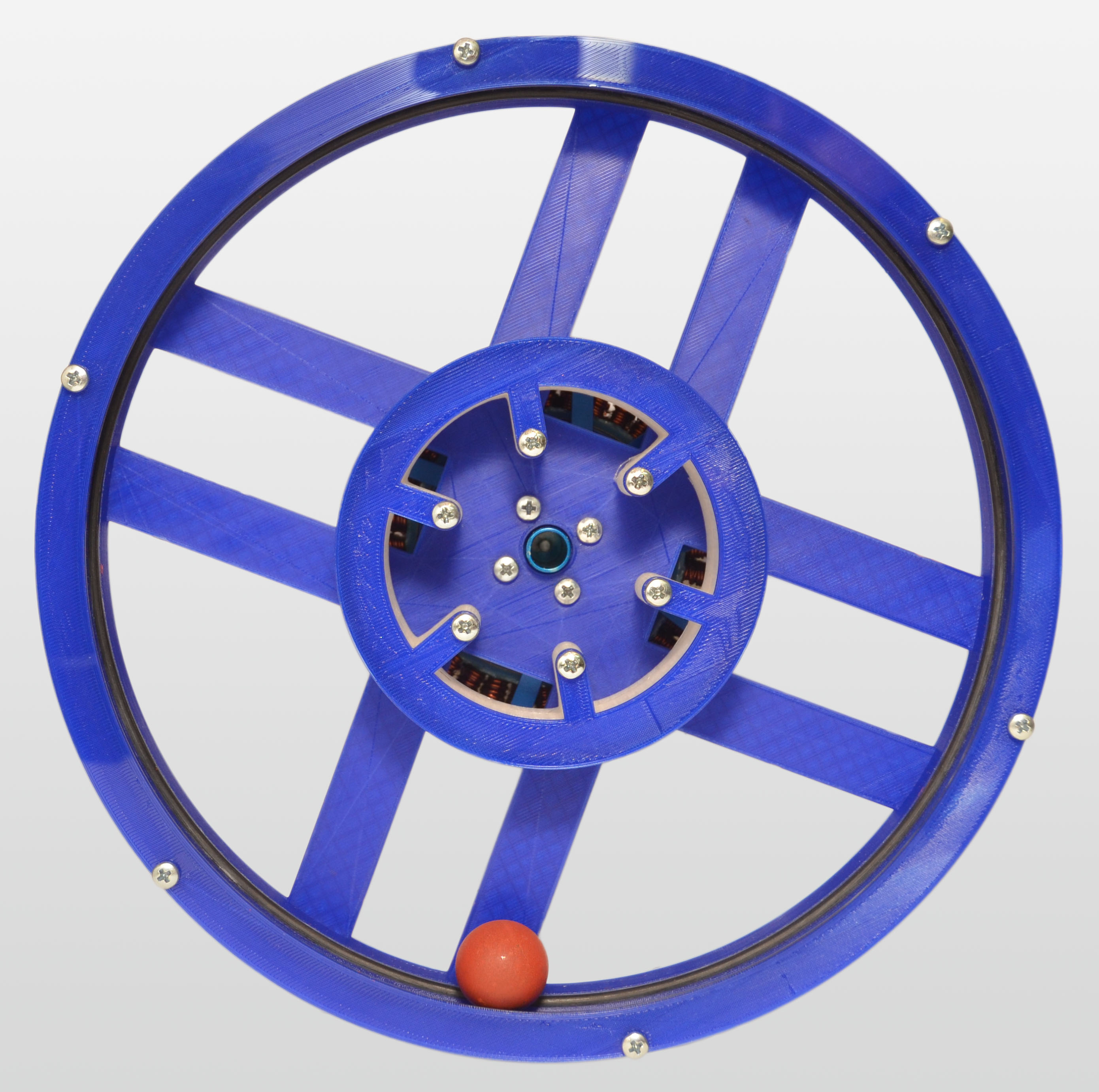}
\caption{A photo of the real hardware setup.}
\label{fig:photo}
\end{center}
\end{figure}

The hardware setup used for the verification is displayed in Fig~\ref{fig:photo}. Parameters of the setup are summarized in Table~\ref{tb:demmodel_params}. The hoop is 3D-printed\footnote{The drawings and codes are available at \url{http://github.com/aa4cc/flying-ball-in-hoop}} and it has imbedded rubber o-rings serving as high-friction rails for a metal ball. The hoop is attached to a BLDC motor driven by a \textit{PearControl ESC3} regulator enabling us to control the motor in a speed and current mode. To simplify the control algorithm, we let the current control to the motor regulator and used the speed mode to set directly a desired angular velocity $\dot{\theta}_\mathrm{des}$. Nevertheless, in Section~\ref{sec:model} we considered $\ddot{\theta}$ to be the control input to the model, thus also the output of the regulator. Therefore, to get $\dot{\theta}_\mathrm{des}$, we numerically integrate the output of the regulator $\ddot{\theta}$. Judging from the experiments, the control system performs surprisingly well in spite of this simplification.

\begin{figure}
    \centering
    \begin{subfigure}[b]{\columnwidth}
        \includegraphics[width=\textwidth]{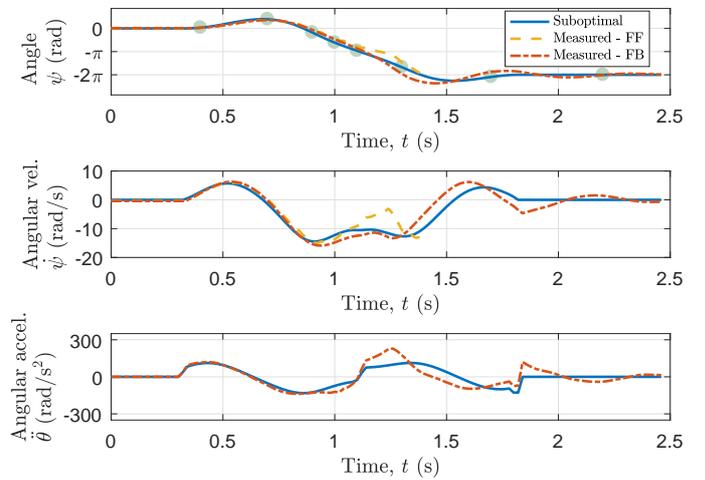}
        \caption{Graphs}
        \vspace{5pt}
        \label{fig:rollIt_exp_graphs}
    \end{subfigure}
    \begin{subfigure}[b]{0.8\columnwidth}
        \includegraphics[width=\textwidth]{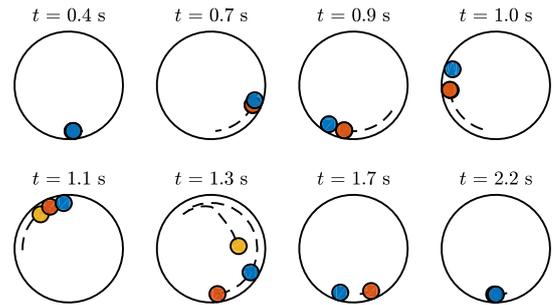}
        \caption{Time instants}
        \label{fig:rollIt_exp_visu}
    \end{subfigure}
    \caption{A comparison of a nominal trajectory for Task 1 with a trajectory measured on the real system with feedback (FB) and without feedback (FF). The comparison has the form of (a) graphs and (b) visualizations of the position of the ball at several time instants. The time instants are also displayed in (a) by green dots. The colors of the balls in (b) correspond to the colors of trajectories in (a). The trajectory without feedback ends up at approximately $t=\SI{1.2}{\second}$ because at that time the ball drops out of the hoop. The dimensions in (b) are not in scale.}
    \label{fig:rollIt_exp}
\end{figure}


\subsection{Implementation} 
\label{sub:implementation}
The control law was discretized by sampling of~\eqref{eq:ctrl_law} and it was implemented in \textit{Simulink}. We used \textit{Support Package for Raspberry Pi Hardware} to generate code, compile it, and run it on \textit{Raspberry Pi 3}. Raspberry Pi 3 has sufficient computational power that it also allows us to measure the position of the ball by processing images from \textit{Raspberry Pi camera module}. The image-processing algorithm was implemented in \textit{Python} and \textit{OpenCV}. Both, the image processing and control algorithm run at \SI{50}{\hertz}.

\begin{figure}
    \centering
    \begin{subfigure}[b]{\columnwidth}
        \includegraphics[width=\textwidth]{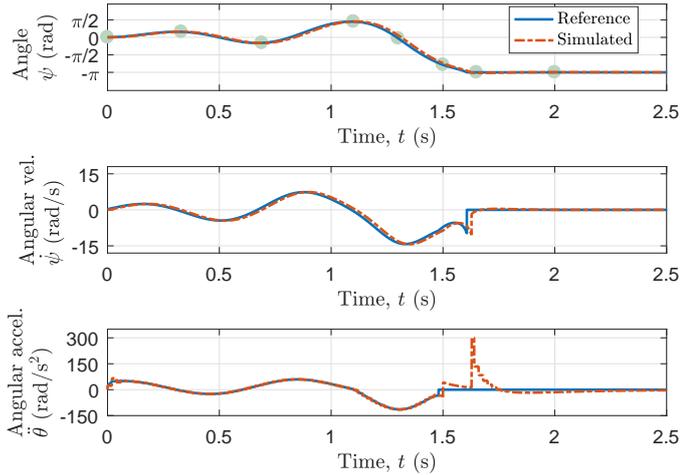}
        \caption{Graphs}
        \vspace{5pt}
        \label{fig:getItUp_exp_graphs}
    \end{subfigure}
    \begin{subfigure}[b]{0.8\columnwidth}
        \includegraphics[width=\textwidth]{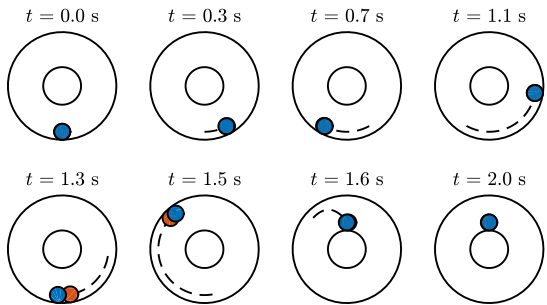}
        \caption{Time instants}
        \label{fig:getItUp_exp_visu}
    \end{subfigure}
    \caption{A comparison of a nominal trajectory for Task 2 with a trajectory obtained by simulation of the model with the trajectory stabilizing regulator. The comparison has the form of (a) graphs and (b) visualizations of the position of the ball at several time instants. The time instants are also displayed in (a) by green dots and the colors of the balls in (b) correspond to the colors of trajectories in (a). The dimensions in (b) are not in scale.}
    \label{fig:getItUp_exp}
\end{figure}

The trajectory stabilizing regulator, as a state regulator, needs to know values of all states. States $x_1=\theta$ and $x_2=\dot{\theta}$ do not need to be measured because they are given by integration of the input $u=\ddot{\theta}$. Angle of the ball $\psi$ is measured by a camera and image processing. Nevertheless, the angular velocity $\dot{\psi}$ is rather difficult to measure directly and thus has to be estimated. For this purpose, we used a discrete \textit{Extended Kalman Filter (EKF)} in the standard setting.

Beside the model mismatch and state estimation, in the reality, one also usually has to deal with a latency in the measurement. In our hardware setup, we measured the latency of the measurement of $\psi_\mathrm{meas}$ to be approximately two control periods (\SI{40}{\milli\second}). Thus, at control period $k$, we at first estimate the state $\hat{\mathbf{x}}[k-2]$ by EKF based on the previous estimate $\hat{\mathbf{x}}[k-3]$, $\psi_\mathrm{meas}[k]$ and $u[k-3]$ and then compensate for the delay by two-step prediction:
\begin{align}
  \hat{\mathbf{x}}[k-1] &= \hat{\mathbf{x}}[k-2] + T f_\mathrm{S1}(\hat{\mathbf{x}}[k-2], u[k-2]), \\
  \hat{\mathbf{x}}[k] &= \hat{\mathbf{x}}[k-1] + T f_\mathrm{S1}(\hat{\mathbf{x}}[k-1], u[k-1]),
\end{align}
where $T$ is the control period and it equals to $\SI{20}{\milli\second}$.

\subsection{Verification} 
\label{sub:verfication}
\vspace{-3pt}
Regarding Task 1, a reference trajectory $\mathbf{x}^\ast$ obtained by the solution of~\eqref{eq:rollIt_OCP} together with a trajectory measured on the real hardware setup are shown in Fig.~\ref{fig:rollIt_exp}. Apparently, the validity of the control algorithm is verified; it is able to follow the desired trajectory to such an extent that the ball ends up in the desired position and does not drop out of the hoop. We have not yet implemented the solution of Task 2 on the real hardware setup. Nevertheless, the proposed solution works very nicely in simulations as Fig.~\ref{fig:getItUp_exp} shows. Solutions of both tasks are also presented in a video clip available at~\url{http://youtu.be/GBKhRHtjpvQ}.

\vspace{-1pt}

\section{Conclusion} 
\label{sec:conclusion}
\vspace{-1pt}
We showed that the well-known ball and hoop system can be used not only to demonstrate some aspects of linear control theory but also to demonstrate strengths of numerical optimal control. We extended the ball and hoop system by another hoop and described this extended system by a hybrid model. We presented two tasks for the extended system that are ideal for a demonstration of trajectory generation and stabilization and solve them. As the model of the extended system is simple, one can also use it for a demonstration of control techniques that are based on \textit{Sum-Of-Squares} programming. For instance, our proposed solution of the presented tasks can be extended by \textit{LQR trees} (\cite{Tedrake2010LqrTrees}) to work also for varying initial conditions. Furthermore, the inner hoop can be replaced by something else. For example, one can replace it by a beam and modify the Task 2 to ``get the ball on a beam and stabilize it there''. Or, the inner hoop can be replaced by a figure-eight shaped rails and the ``butterfly'' task can be solved on the system~(\cite{Cefalo2006EnergyBased}). 
\bibliography{Remote}

\end{document}